\documentclass[doublecol, figures]{epl2} 

\usepackage{latexsym}
\usepackage{amsmath}
\usepackage{amsbsy}

\title{Spin-Hall effect theory: new analytical solutions of the Pauli equation in a quantum dot}
\shorttitle{Spin-Hall effect theory: new analytical solutions of the Pauli equation  \dots}

\author{J. L. Cardoso}
\shortauthor{J. L. Cardoso}

\institute{                    
Area de  F\'{\i}sica Te\'orica y Materia Condensada, UAM-Azcapotzalco,
Avenida San Pablo 180, C\'odigo Postal 02200 M\'exico Distrito Federal,
M\'exico
}

\abstract{
In this work, we present the analytical solution of the effective mass Pauli equation, with Rashba and linear Dresselhaus interactions, for an electron gas moving through a semiconductor quantum dot under a longitudinal electric field, which is defined along the $x$-direction. We study the relative influence of the Rashba and Dresselhaus terms on the spin-Hall effect for the first propagating and edge channels, by analyzing the mixing between spin-up and -down states and the zero-field spin splitting along the transverse directions. When the spin rotation depends only on the $y$-coordinate, the spin orientation and the spin density vary along this transverse coordinate and, in this case, we show that the spin-Hall effect is only due to the Dresselhaus term, for depolarized electrons. On the other hand, if the spin rotation depends on the $z$-coordinate, the spin-Hall effect is provoked only by the Rashba interaction.
}

\begin{document}

\maketitle

\section{Introduction}
In a semiconductor, a two-dimensional electron gas (2DEG) in the conduction band does not experience as strong a nuclear attraction as an electron in an atom does. However, it may still see electric fields due to internal effects. Those symmetry-breaking electric fields, on one hand, can induce a spin splitting due to a structural inversion asymmetry, which is named Rashba spin-orbit interaction\cite{EIRashba, SDatta}. On the other hand, the well-known Dresselhaus spin-orbit interaction\cite{GDresselhaus} can induce a splitting by electric fields that arise due to bulk (or crystallographic) inversion asymmetry. Externally applied electric fields may also provoke such spin splittings\cite{JEHirsch, SZhang, RdeSousa, SPrabhakar1, SPrabhakar2}. The full understanding of the spin dynamics in a semiconductor heterostructure is of great interest from the theoretical and the applied physics points of view\cite{IZutic}. Taking into account both spin-orbit interactions (SOI), theoretical work has been focused on the investigation of electron transport in a 2DEG \cite{JEHirsch, SZhang, RdeSousa, SPrabhakar1, SPrabhakar2, IZutic, WXu, MValin-Rodriguez, NHatano, MDuckheim, JBruning, S-HChen, J-WRhin, ELipparini, MZarea, LJiang, VMRamaglia, DZhang}. We propose here to solve analytically the quasi two-dimensional Pauli eq. with a longitudinal electric field, which induces both SOI. 

Among other experiments, in a 2DEG in strong magnetic fields, the integer Quantum Hall effect and the Hall resistance quantization\cite{KVKlitzing} lead to transverse modes and edge states. In this case, each edge state is spin-degenerate because the Zeeman splitting is negligible\cite{MTorres, AKunold}. Recently, theoreticians predict a different kind of edge channels that can form even in the absence of an external magnetic field\cite{JEHirsch, SZhang, S-QShen, NASinitsyn, SMurakami, CLKane, BABernevig1, BABernevig2}, this anomalous behavior is called spin-Hall effect (SHE). There are experimental demonstrations of the SHE in quantum wells of $\chem{HgTe}$ \cite{MKonig} and of $\chem{InAs/GaSb}$ \cite{IKnez}.

In this work, we study the dynamics of electrons in quasi two-dimensional quantum dots in the presence of Rashba and Dresselhaus spin-orbit interactions, which are provoked by an applied electric field. In this approach (see the next section), we neglect the cubic Dresselhaus spin-orbit term. Here, we work with regions where there is no applied magnetic field. We place special emphasis on how to use the ``variable separation method'' since we have written the eigenspinors as the product of a transverse space dependent part (a particle eigenfunction in an infinite potential well) and a spin dependent part (a rotation operator times another spinor) for each transverse direction. The solution of the longitudinal part is given by the well-known Airy functions. We find spin accumulations near the semiconductor quantum dot edges: spins of one polarization pile up at one edge, while the spins with opposite polarization pile up at the other edge. In other words, it is possible to predict the spin edge states and, therefore, to give a description of SHE. 

\section{The Pauli equation and its solutions}

To study transport properties of a quasi-2DEG in a longitudinal electric field, we shall consider a quasi two-dimensional quantum dot of dimensions $w_x \geq w_y > w_z$, such that $\frac{\pi^2}{w_x^2} \leq \frac{\pi^2}{w_y^2} < \frac{\pi^2}{w_z^2}$. We will solve the effective mass Pauli eq. for an electron moving in the direction $x$ under a longitudinal electric field $\vect{F} =\mathcal{F} \widehat{\imath}$ 
\begin{equation} 
\begin{array}{c}
\left[ \frac{1}{2m} p^2 -e\mathcal{F}x +V(x,y,z) \right. \\
\left. +\tens{H}_{\mathrm R} + \tens{H}_{\mathrm D} -E_{\mathrm F} \right] \vect{\Phi} \left(x,y,z\right) = 0,
\end{array}
\end{equation}
where $\vect{\Phi} \left(x,y,z\right)$ is a spinor, $\vect{p} = -i\hbar \vect{\nabla}$ is the momentum, $m$ is the effective mass, $E_{\mathrm F}$ is the Fermi energy, $V(x,y,z)$ is the confining hard wall potential and $\tens{H}_{\mathrm R}$ and $\tens{H}_{\mathrm D}$ are the Rashba and the Dresselhaus interactions respectively. We consider that both SOI are induced by the applied electric field $\mathcal{F}$ and, therefore, $\tens{H}_{\mathrm R}$ has the following form $\tens{H}_{\mathrm R} = \frac{\alpha^*}{\hbar} \left( \tens{\sigma}_y p_z - \tens{\sigma}_z p_y \right)$, with $\alpha^* =a e\mathcal{F}$ and $a$ the material constant. On the other hand, it is possible to show that the average momentum squared in ground state is given by $\left\langle p_x^2 \right\rangle = 0.7794 \left( \hbar/l_E \right)^2$, where $l_E^3 = \frac{\hbar^2}{2me \mathcal{F}}$ is the electric length \cite{RdeSousa}. Notice that for high values of the electric field, such that $w_z > l_E$, the following inequalities $\left\langle p_x^2 \right\rangle > \left\langle p_z^2 \right\rangle > \left\langle p_y^2 \right\rangle$ can be true and we could propose that the most important term of the linear Dresselhaus interaction is written by $\tens{H}_{\mathrm D} = \frac{\beta^*}{\hbar} \left( \tens{\sigma}_y p_y - \tens{\sigma}_z p_z \right)$, with $\beta^* \simeq 0.7794 \beta_{3D}/l_E^2$ and $\beta_{3D}$ is the Dresselhaus three dimensional term. 

The Pauli eq. under our approach is reduced to the following form
\begin{equation} \label{Sch-SOI}
\begin{array}{c}
\left[ \nabla^2 +\frac{2m E_{\textrm F}}{\hbar^2} +\frac{x}{l_E^3} + 2i\left( \beta \tens{\sigma}_y - \alpha \tens{\sigma}_z \right) \frac{\partial}{\partial y} \right. \\
\left. +2i \left(  \alpha \tens{\sigma}_y - \beta \tens{\sigma}_z \right) \frac{\partial}{\partial z} \right] \vect{\Phi} \left(x,y,z\right) = 0,
\end{array}
\end{equation}
here $\alpha = m \alpha^*/\hbar^2$ and $\beta = m \beta^*/\hbar^2$. In order to get rid of the variable $z$, we will consider the complete set of stationary states of a particle in a $z$-dimensional infinite potential well with boundary conditions $Z_n (-w_z/2) = Z_n (w_z/2) = 0$, which are solutions of the differential eq.
\begin{equation} \nonumber
\begin{array}{c}
\frac{d^2 \vect{\zeta}}{dz^2} + 2i \left( \alpha \tens{\sigma}_y - \beta \tens{\sigma}_z \right) \frac{d \vect{\zeta}}{dz} + q_z \vect{\zeta} =0 .
\end{array}
\end{equation}
It is easy to verify that $\vect{\zeta} = \sum_n \vect{\zeta}_n = \tens{R}_z \sum_n Z_n \left(z \right) \vect{c}_n$, where $\vect{\zeta}$, $\vect{\zeta}_n$ and $\vect{c}_n$ are spinors,
\begin{equation} \nonumber
 Z_n \left(z \right) = \left\{
\begin{array}{cl}
\cos \left( k_{z,n} \ z\right) & \textrm{ for } n \textrm{ odd } \\
\sin \left( k_{z,n} \ z\right) & \textrm{ for } n \textrm{ even }
\end{array} ,
\right.
\end{equation}
$k_{z,n} = \frac{n \pi}{w_z}$, $q_z = k_{z,n}^2 +\alpha^2 +\beta^2$ and $\tens{R}_z = e^{-i \left( \alpha \tens{\sigma}_y - \beta \tens{\sigma}_z \right)z}$ is a rotation operator. If we use the complete set of functions $\left\{ \vect{\zeta}_n \right\}$ to expand $\vect{\Phi} \left(x,y,z\right)$, we get
\begin{equation} \label{ansatz}
\begin{array}{c}
\vect{\Phi} \left(x,y,z\right) = \tens{R}_z \sum_{n=1}^{\infty} Z_n (z) \vect{F}_n (x,y).
\end{array}
\end{equation}
By introducing this function in the Pauli eq. (\ref{Sch-SOI}), multiplying by $\tens{R}_z^\dagger Z_l \left(z \right)$ and integrating on the variable $z$, we have
\begin{equation}
\begin{array}{c}
\left\{ \frac{\partial^2}{\partial x^2} +\frac{\partial^2}{\partial y^2} +\frac{2m E_{\textrm F}}{\hbar^2} -k_{z,l}^2 +\frac{x}{l_E^3} +\eta^2 \right. \\
+ \left[ \frac{\alpha^2-\beta^2}{\eta}  \tens{\sigma}_x I_{l,l} +\frac{\beta \tens{\sigma}_y}{\eta^2} \left( 2\alpha^2 - \left( \alpha^2 -\beta^2 \right) J_{l,l} \right) \right. \\
\left. \left. -\frac{\alpha \tens{\sigma}_z}{\eta^2} \left( 2\beta^2 + \left( \alpha^2 -\beta^2 \right) J_{l,l} \right) \right] \frac{\partial}{\partial y} \right\} \vect{F}_l \\
+\frac{\alpha^2 -\beta^2}{\eta^2} \sum_{n} \left[ \eta \tens{\sigma}_x I_{n,l} -\left( \beta \tens{\sigma}_y -\alpha \tens{\sigma}_z \right) J_{n,l} \right] \frac{\partial \vect{F}_n}{\partial y} =0
\end{array}
\end{equation}
here $\eta = \sqrt{\alpha^2 +\beta^2}$ and the mixing terms are given by $I_{n,l} = \frac{2}{w_z} \int_{-w_z/2}^{w_z/2} \sin 2\eta z \ Z_n (z) Z_l (z) dz$ and $J_{n,l} = \frac{2}{w_z} \int_{-w_z/2}^{w_z/2} \cos 2\eta z \ Z_n (z) Z_l (z) dz$. Notice that if $\beta = \pm \alpha$ the previous eqs. are uncoupled.

When the width $w_z$ is narrow, the energy levels spacing, that is defined by $\Delta E_{z,l} = \frac{\hbar^2}{2m} \left( k_{z,l+1}^2 -k_{z,l}^2 \right) = \frac{\pi^2 \hbar^2}{2mw_z} \left( 2l+1 \right)$, is wide. Therefore, the transverse resonances are clearly separated in energy in the quasi two-dimensional quantum dot and $\eta z \leq \eta w_z << 1$. We are able to take the approximations $\sin 2\eta z \simeq 2\eta z$ and $\cos 2\eta z \simeq 1$, in such case $I_{1,1} \simeq 0$, $J_{1,1} \simeq 1$, $I_{1,2} \simeq \frac{16 \eta w_z}{9 \pi^2}$ and  $J_{1,2} \simeq 0$. In the case where there is only one propagating mode (for Fermi energies $E_{\textrm F} < \Delta E_1$), we are left with the 2D differential eq.
\begin{equation}\label{spinor}
\begin{array}{c}
\frac{\partial^2 \vect{F}_1} {\partial x^2} + \frac{\partial^2 \vect{F}_1} {\partial y^2} + \left[ \frac{2m}{\hbar^2}E_{\mathrm F} - \frac{\pi^2}{w_z^2} + \eta^2 + \frac{x}{l_E^3} \right] \vect{F}_1 \\
 + 2i \left( \beta \tens{\sigma}_y - \alpha \tens{\sigma}_z \right)  \frac{\partial \vect{F}_1}{\partial y} = 0,
\end{array}
\end{equation}
which describes the quantum dot in the 2DEG limit. We neglect here the term $I_{1,1}$, because $\left| \frac{\alpha^2 -\beta^2}{\eta} \right| \frac{16 w_z}{9 \pi^2} << 1$. For clarity, the index $n = 1$ shall be suppressed in the following expressions. Taking into account the ansatz given by (\ref{ansatz}), we use a similar one
\begin{equation}\label{ansatz:2}
\begin{array}{c}
\vect{F} \left( x,y \right) = \tens{R}_y \vect{S} \sum_{l=1}^\infty Y_l (y) X_{l} (x)
\end{array}
\end{equation}
where $\tens{R}_y = e^{-i \left( \beta \tens{\sigma}_y - \alpha \tens{\sigma}_z \right)y}$ is another rotation operator acting on the $yz$ plane, $X_l$ is a scalar function, $Y_l (y)$ is an eigenstate of a particle in an infinite potential well and $\vect{S}$ is a spinor, which is independent of $x$. The sum runs over the $N$ propagating modes, with transverse wave number $\kappa_{\perp,l}^2 = \left( l \pi/w_y \right)^2 +\left( \pi/w_z \right)^2$ such that $\kappa_N^2 = \frac{2 m}{\hbar^2} E_{F} - \kappa_{\perp,l}^2 \geq 0$, and an infinite number of evanescent modes. For each longitudinal wavenumber $\kappa_{l}$ there are two propagating physical channels: one with spin-up and another with spin-down. Equation (\ref{spinor}) gets transformed into the following system of uncoupled eqs.
\begin{equation} \label{longitudinal}
\begin{array}{c}
\frac{\upd^2 X_{l}}{\upd x^2} + \left[ \kappa_l^2 +2\eta^2 +\frac{x}{l_E^3} \right] X_{l} =0
\end{array}
\end{equation}
for each Fourier open channel $l$ at the $y$ direction. The general solutions of these eqs. are given by 
\begin{equation} \label{solut:F}
\begin{array}{c}
X_{l} \left(x \right) = \textrm{Ai} \left(-\chi_l \right) C_0 + \textrm{Bi} \left(-\chi_l \right) C_1 ,
\end{array}
\end{equation}
being
\begin{equation} 
 \begin{array}{c}
\textrm{Ai} \left( -\chi_l \right) = \frac{\sqrt{\chi_l}}{3} \left[ J_{-\frac{1}{3}} \left( \frac{2}{3}\chi_l^{\frac{3}{2}} \right)+ J_{\frac{1}{3}} \left( \frac{2}{3}\chi_l^{\frac{3}{2}} \right) \right]\\
\textrm{Bi} \left( -\chi_l \right) = \sqrt{\frac{\chi_l}{3}} \left[ J_{-\frac{1}{3}} \left( \frac{2}{3}\chi_l^{\frac{3}{2}} \right)- J_{\frac{1}{3}} \left( \frac{2}{3}\chi_l^{\frac{3}{2}} \right) \right] 
\end{array}
\end{equation}
the well-know Airy functions and $\chi_l = \frac{x}{l_E} + \left( \kappa_l^2 + 2\eta^2 \right) l_E^2$. If we follow the procedure developed in \cite{RdeSousa}, the solution (\ref{solut:F}) is reduced and it has the following form
\begin{equation} \nonumber
 \begin{array}{c}
X_{1,l} \left(x \right) = 1.4261 l_E^{-1/2} \textrm{Ai} \left( \frac{x}{l_E}+r_1  \right)
\end{array}
\end{equation}
with $r_1 = -2.3381$ as the first zero of $\textrm{Ai}$ and $E_{x,1} = -r_1 e\mathcal{F}l_E$ as the ground state energy. Here, the Fermi energy is totally quantized $E_{\textrm F, 1,l} = E_{x,1} + \frac{\hbar^2}{2m} \kappa_{\perp,l}^2$. From now on, we will suppress the index 1 for all expressions. Notice that those eqs. indicate that there is no mixing among the $l$-Fourier channels. Due to the rotation operator $\tens{R}_y$, there is mixing between the spin channels. 

\section{Edge states as the origin of the spin-Hall effect} 
The spin-Hall effect is due to spin accumulations near the semiconductor quantum-dot edges even in the absence of an external magnetic field. In this section, we will show that depending on the Dresselhaus interaction, these rotations might provoke spin accumulations close to $y = -\frac{w_y}{2}$ or to $y = \frac{w_y}{2}$ for electronic depolarized currents.  We shall analyze the principal characteristics of the spin-transverse wave functions at a particular case where there are depolarized electrons. 

According to the previous section, the spin-dependent transverse wave function is written by
\begin{equation}\label{solut:planeF}
\begin{array}{c}
\vect{F}_{l} \left( x,y \right) = \tens{R}_y \vect{S} Y_l \left( y \right) X_{l} \left( x \right)
\end{array}
\end{equation}
The exponential term $\tens{R}_y = e^{-i \left( \beta \tens{\sigma}_y - \alpha \tens{\sigma}_z \right)y}$ is a rotation operator and indicates a precession about an axis in the plane $yz$. Due to the Rashba interaction, electron spins rotate around the $z$ axis, while the Dresselhaus term causes rotation around the $y$ axis. The mixing between spin channels depends only on the strength of the Dresselhaus term. 

\begin{figure}
\centerline{\includegraphics[angle=0,width=3.0in]{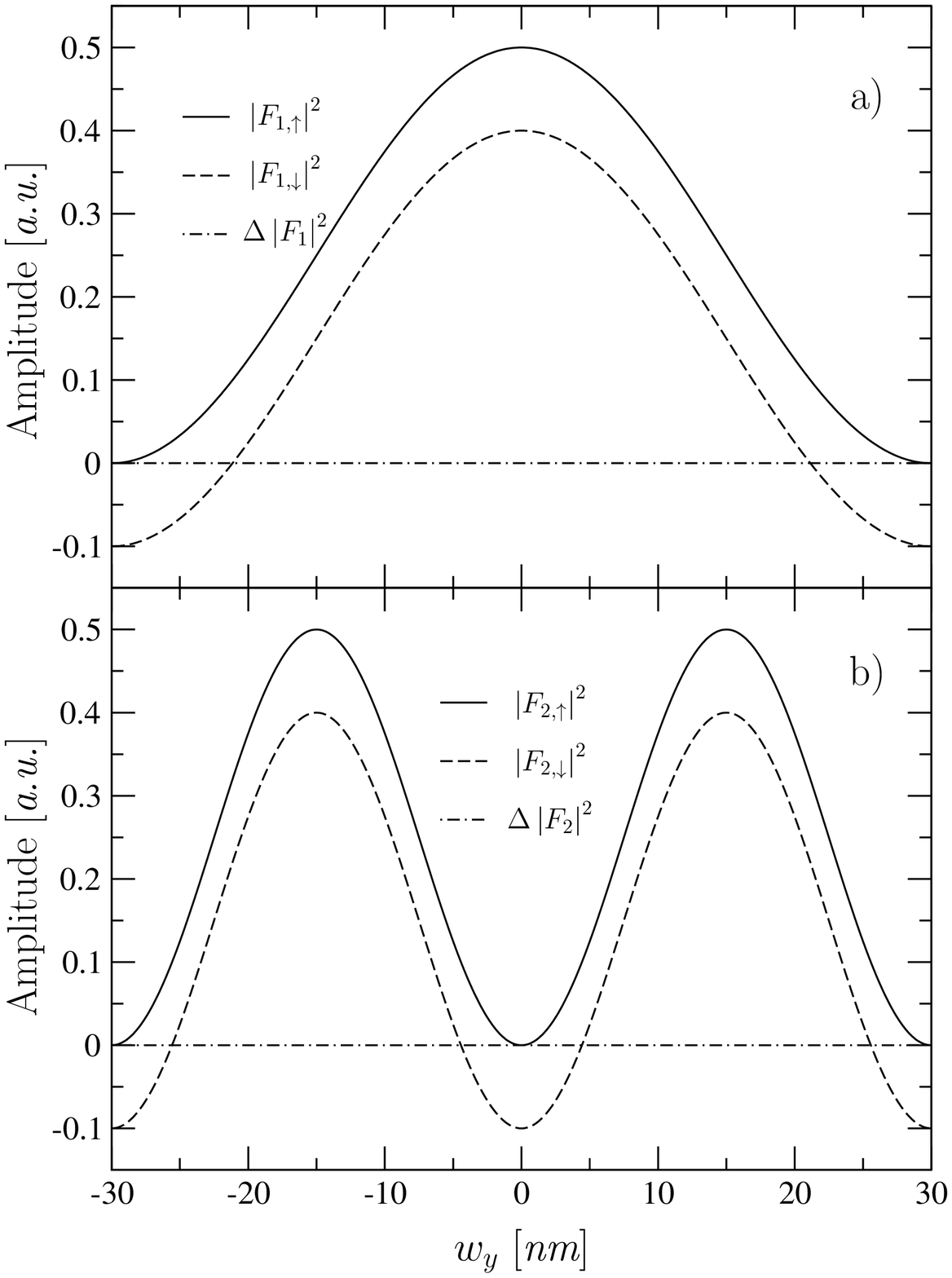}} 
\caption{ $\left| F_{l,\uparrow} \right|^2$, $\left|  F_{l,\downarrow} \right|^2$ and $\Delta \left|  F_{l} \right|^2$ as functions of the $y$ direction when there is no Dresselhaus term $\beta = 0$ and $\alpha \neq 0$, for spin depolarized electrons at the $z$ direction: a) when $l = 1$; and b) when $l = 2$. For the sake of clarity, the curves for $\left| F_{l, \uparrow} \right|^2$ and $\left| F_{l, \downarrow} \right|^2$ are vertically offset by $0.1$ amplitude units. The electron spins only precede about $z$ direction, preserving their own initial polarization, and this motivates that one can not find any spin accumulation near the edges.}
\label{SHE-R}
\end{figure}

\begin{figure}
\centerline{\includegraphics[angle=0,width=3.0in]{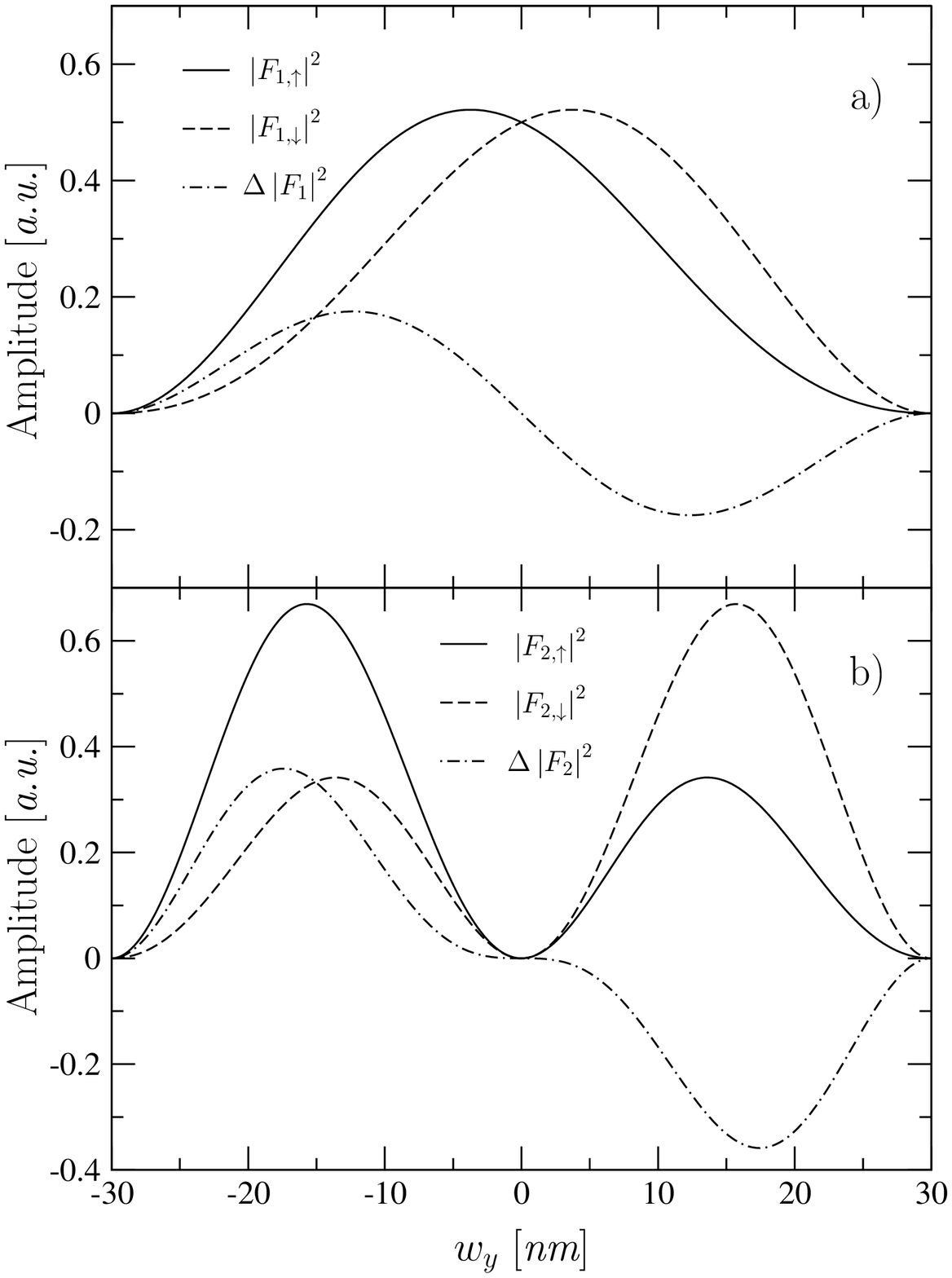}}
\caption{ $\left| F_{l,\uparrow} \right|^2$, $\left|  F_{l,\downarrow} \right|^2$ and $\Delta \left| F_l \right|^2$ as functions of the $y$ direction when here is no Rashba term, $\alpha = 0$ and $\beta \neq 0$ for spin depolarized electrons at the $z$-direction: a) when $l = 1$; and b) when $l = 2$. The symmetry of $\left| F_{l,\uparrow} \right|^2$ and $\left|  F_{l,\downarrow} \right|^2$ is broken by the Dresselhaus term. The spin-up electrons pile up at the left edge because $\Delta \left| F_{l} \right|^2  > 0$, while $\Delta \left| F_{l} \right|^2  < 0$ indicating that the spin-down electrons pile up at the right edge.}
\label{SHE-D}
\end{figure}

It is well known that the $\chem{GaSb}$ has the following characteristics: $a = 33 \un{\AA^2}$, $\beta_{3D} = 187 \un{eV \AA{}^3}$ and $\frac{m}{m_0} = 0.0412$ \cite{RdeSousa}. On the other hand, we use the following parameters: $\mathcal{F} = 5.0 \times 10^5 \un{\frac{eV}{cm}}$, $w_x = w_y = 60 \un{nm}$ and $w_z =20 \un{nm}$ for the first longitudinal ground-state energy $E_{x,1} = -r_1 e \mathcal{F} l_E = 0.3345\un{eV}$. In figs. \ref{SHE-R} and \ref{SHE-D} we plot the spinors' amplitudes $\left| F_{l, \uparrow} \right|^2$, $\left| F_{l,\downarrow} \right|^2$ and the difference between them $\Delta \left| F_{l} \right|^2 = \left| F_{l, \uparrow} \right|^2 - \left| F_{l, \downarrow} \right|^2 $ as functions of the $y$ direction, for $l = 1$ and 2. 

Figure \ref{SHE-R} shows the spin amplitudes for spin depolarized electrons ($S_{l,\uparrow} = 1/\sqrt{2}$ and $S_{l,\downarrow} = 1/\sqrt{2}$) considering that there is no Dresselhaus term. For the sake of clarity, the curves for $\left| F_{l, \uparrow} \right|^2$ and $\left| F_{l, \downarrow} \right|^2$ are vertically offset $0.1$ amplitude units. The electron spins precess only around the $z$ direction, preserving their own initial polarization, because there is only the Rashba interaction, thus one can not find any spin accumulation near the edges. In this way, the Rashba term does not affect the symmetry of $\left| F_{l,\uparrow} \right|^2$ and $\left|  F_{l,\downarrow} \right|^2$, because $\left| F_{l,\uparrow} \right|^2 = \left| F_{l,\downarrow} \right|^2 = Y_l^2 \left( y \right)$ for $l=1,2$. Therefore, there is no difference between the spinor amplitudes: on both cases $\Delta \left| F_l \right|^2 = 0$. 

Let us now cancel the Rashba contribution. Figure \ref{SHE-D} shows that for spin depolarized electrons the symmetry of $\left| F_{l,\uparrow} \right|^2$ and $\left| F_{l,\downarrow} \right|^2$ are broken by the Dresselhaus term. In this case, the exponential term is reduced to the following form 
\begin{equation}  
 e^{-i \tens{\sigma}_y \beta y} = \left[ 
\begin{array}{cc}
\cos \left( \beta y \right) & - \sin \left( \beta y \right) \\
\sin \left( \beta y \right) & \cos \left( \beta y \right)
\end{array} 
\right],
\end{equation}
this exponential term indicates mixing between both spin polarizations because the off-diagonal terms are real and different from zero for processes where a spin flip has taken place in the transverse $y$-direction; and, therefore, the spin amplitudes are given by
\begin{equation}  
 \begin{array}{c}
\left| F_{l,\uparrow} \right|^2 = \frac{1}{2} \left[ 1-\sin \left( 2\beta y \right) \right] Y_l^2 \left( y \right) \\
\left| F_{l,\downarrow} \right|^2 = \frac{1}{2} \left[ 1+\sin \left( 2\beta y \right) \right] Y_l^2 \left( y \right)
\end{array}
\end{equation}
where
\begin{equation}  
 Y_l \left(y \right) = \left\{
\begin{array}{cl}
\cos \left( k_{y,l} y\right) & \textrm{ for } l \textrm{ odd } \\
\sin \left( k_{y,l} y\right) & \textrm{ for } l \textrm{ even }
\end{array}
\right.
\end{equation}
and $k_{y,l} = \frac{l \pi}{w_y}$. The factor corresponding to a stationary particle in a infinite potential well $Y_l^2 \left( y \right)$ dominates at the edges' neighborhood and at its possible nodes, causing suppression of the spin-dependent amplitudes in such zones. In the neighborhood of $y=0$, $\left| F_{l,\updownarrow} \right|^2$ are dominated by the $1\pm \sin \left( 2\beta y \right)$ factors. Both factors are positive and $1- \sin \left( 2\beta y \right)$ increases while $1+ \sin \left( 2\beta y \right)$ decreases when $y$ grows. The maxima of $\left| F_{l,\uparrow} \right|^2$ are moved and grow to the left side by $1- \sin \left( 2\beta y \right)$, while for $\left| F_{l,\downarrow} \right|^2$ maxima are moved to the right side due to the $1+ \sin \left( 2\beta y \right)$ term. We must notice, however, that $\Delta \left| F_l \right|^2 = -\sin \left( 2\beta y \right) Y_l^2 \left( y \right)$ and, therefore, the spin-up electrons pile up at the left edge because $\Delta \left| F_l \right|^2  > 0$, while $\Delta \left| F_l \right|^2  < 0$ indicates that the spin-down electrons pile up at the right edge, these spin accumulations imply a mixing between the spin populations. In such cases, $\left| F_{l, \uparrow} \right|^2$ and $\left| F_{l, \downarrow} \right|^2$ become the well-known edge states.

The behavior is described in figs. \ref{SHE-R} and \ref{SHE-D}, shown only for the $l = 1,2$ states. For other states the general trends are conserved. In this way, we only discussed them for $l =1$ and 2. 

When there are both SOI, the exponential term now is written in the following form
\begin{equation}  
 \begin{array}{l}
e^{-i \left( \beta \tens{\sigma}_y - \alpha \tens{\sigma}_z \right)y} = \\
\left[ \begin{array}{cc}
\cos \left( \eta y \right) + i\frac{\alpha}{\eta} \sin \left( \eta y \right) & - \frac{\beta}{\eta} \sin \left( \eta y \right) \\
\frac{\beta}{\eta} \sin \left( \eta y \right) & \cos \left( \eta y \right) - i\frac{\alpha}{\eta} \sin \left( \eta y \right)
\end{array} \right],
\end{array}
\end{equation}
the off-diagonal matrix elements $\mp \frac{\beta}{\eta} \sin \left( \eta y\right)$ reflect the passage of flux from one spin state to another. The origin of these transitions is the Dresselhaus spin-orbit interaction that stimulates a mixing between the propagating physical channels, in the transverse regions where the $\beta$ is significant. On the other hand, the spin-dependent edge states are given by $\left| F_{l,\uparrow} \right|^2 = \frac{1}{2} \left[ 1-\frac{\beta}{\eta} \sin \left( 2\eta y \right) \right] Y_l^2 \left( y \right)$ and $\left| F_{l,\downarrow} \right|^2 = \frac{1}{2} \left[ 1+\frac{\beta}{\eta} \sin \left( 2\eta y \right) \right] Y_l^2 \left( y \right)$.

The eigen-energies, the spin-polarized wave functions and the zero-field spin splitting on the $y$-direction can be obtained even if we restrict our analysis to energies bellow the $N$ propagating mode threshold $E_{\mathrm F} \leq \hbar^2 \kappa_{N+1}^2/2m = \frac{\hbar^2}{2m} \left[ \left( (N+1) \pi/w_y \right)^2 +\left( \pi/w_z \right)^2 \right]$ and neglect the evanescent mode contributions. In this way, we found the analytic solutions given by eq. (\ref{solut:planeF}). This solution reflects the mixing of spin $\uparrow$ and spin $\downarrow$ states in the  $yz$-plane. The propagating modes of the spin polarized states are obtained after a unitary transformation
\begin{equation}  
 \begin{array}{c}
\tens{U}: \vect{F}_l \to \widetilde{\vect{F}}_l = \tens{U} e^{-i \left( \beta \tens{\sigma}_y - \alpha \tens{\sigma}_z \right)y} \tens{U}^\dagger \tens{U} \vect{S} Y_l X_l \left(x \right) \\
\widetilde{\vect{F}}_l \left(x,y \right)  = e^{i \tens{\sigma}_z \eta y} \widetilde{\vect{S}} Y_l \left(y \right) X_l \left(x \right) .
\end{array}
\end{equation}
Here,
 \begin{equation}  \nonumber
 \begin{array}{c}
\tens{U} = \tens{U}^\dagger = \frac{1}{\sqrt{2 \eta \left( \alpha + \eta \right)}} \left[ \left( \alpha+\eta \right) \tens{\sigma}_z - \beta \tens{\sigma}_y \right]
\end{array}
\end{equation}
is the relationship of the similarity matrix, $\widetilde{\vect{F}}_l \left(x,y \right) = \tens{U} \vect{F}_l \left(x,y \right)$ and $\widetilde{\vect{S}} = \tens{U} \vect{S}$. In this representation for the right-side propagating state spinors at any point $y$ inside the wave guide, the energies of the spin $\uparrow$ and spin $\downarrow$ states are $E_\uparrow = \frac{\hbar^2}{2m} \left[ k_{y,r}^2+\eta^2 - 2 k_{y,r} \eta \right]$ and $E_\downarrow = \frac{\hbar^2}{2m} \left[ k_{y,r}^2+\eta^2 + 2 k_{y,r} \eta \right]$ and the zero-field spin splitting is given by $E_\downarrow - E_\uparrow = \frac{2 \hbar^2}{m} k_{y,r} \eta$. 

On the other hand, taking into account the 2D effective mass Pauli eq. (\ref{spinor}) we can define an {\it electric current density} $\mbox{\boldmath{$j$}}$ that guarantees the conservation of probability. The $y$ component of such current is given by
\begin{equation} \nonumber
\begin{array}{c}
j_{y,l} = \frac{-ie\hbar}{2m} \left[ \frac{\partial \vect{F}_l^{\dagger}}{\partial y} \vect{F}_l - \vect{F}_l^{\dagger} \frac{\partial \vect{F}_l}{\partial y} -2i \vect{F}_l^\dagger \left( \beta \tens{\sigma}_y -\alpha \tens{\sigma}_z  \right) \vect{F}_l \right]
\end{array}
\end{equation}
with
\begin{equation}  
\begin{array}{c}
 \vect{F}_l \left( x,y \right) = \tens{R}_y \vect{S} Y_l \left(y \right) X_l \left(x \right) 
\end{array}
\end{equation}
the complete analytic solution. Notice that $j_{y,l} = 0 $ for $\vect{F}_l$. This means that, at variance with the classical Hall effect, no Hall voltage builds up between the two edges since no charge imbalance results from this phenomena at the $y$-transverse direction. This spin imbalance is part of the spin texturing and forms an anomalous Hall effect, the spin-Hall effect. However, given that quantum dot is made of a non-magnetic semiconductor, it has equal populations of electrons with spin-$\uparrow$ and -$\downarrow$. 

To solve eq. (\ref{Sch-SOI}) we could follow a different procedure that leads to a differential eq. in only one variable $x$ similar to (\ref{longitudinal}). Let us suppose that the quasi-2DEG is in the $xz$-plane. Now $w_x \geq w_z > w_y$. To get rid of the variable $y$, we will consider the complete set of stationary states of a particle in a $y$-dimensional infinite potential well and, following similar steps to get (\ref{spinor}), we obtain
\begin{equation}
\begin{array}{c}
\frac{\partial^2 \vect{G}_1} {\partial x^2} + \frac{\partial^2 \vect{G}_1} {\partial y^2} + \left[ \frac{2m}{\hbar^2}E_{\mathrm F} - \frac{\pi^2}{w_z^2} + \frac{x}{l_E^3} \right] \vect{G}_1 \\
 + 2i \left( \alpha \tens{\sigma}_y - \beta \tens{\sigma}_z \right)  \frac{\partial \vect{G}_1}{\partial y} = 0,
\end{array}
\end{equation}
Taking into account the ansatz given by (\ref{ansatz:2}), we use a similar one
\begin{equation}
\begin{array}{c}
 \vect{G} \left(x,z\right) = e^{-i \left( \alpha \tens{\sigma}_y - \beta \tens{\sigma}_z \right)z} \vect{S} \sum_{l=1}^{\infty} Z_{l} (z) X_l (x),
\end{array}
\end{equation}
and so on. It is clear that, in this case, the Rashba interaction provokes edge states at the $z$-direction. 

\section{Summary and conclusions}

We study the transport properties of spin-$\frac{1}{2}$ electrons in an homogeneous semiconductor quantum dot subject to an externally longitudinal electric field. By describing the interacting evolution of these electrons with Rashba and linear Dresselhaus interactions, we found the analytic solution of the Pauli eq. for spin-up and spin-down coupled channels and determined the spin texturing; in other words, the spin orientation and the spin density can vary along the transversal directions. On the basis of this phenomena, we found the edge states. Such solutions are in terms of a rotation operator and it indicated a precession about an axis in the plane $yz$, rotations that might provoke spin accumulations close to the edges. There are no charge imbalance resulting from this phenomena at the $y$-transverse direction, because the transverse current is zero and, therefore, there is no Hall voltage build-up between the two edges. 

We can propose, alternatively, that the slab which defines the quantum dot is in the $xz$-plane, here $w_x \geq w_z >> w_y$. The 2D effective mass Pauli eq. now is in terms of the transverse $z$-direction. When we solve it, taking into account the rotation operator $e^{-i \left( \alpha \tens{\sigma}_y - \beta \tens{\sigma}_z \right)z}$, the edge states and the mixing between both spin polarizations are provoked only by the Rashba term.

\acknowledgments
The author would like to thank Professors H. Hern\'andez-Salda\~na, A. Kunold, and J. Grabinsky for clarifying discussions. 

\bibliographystyle{eplbib}
\bibliography{SHE-extrinsic}

\end{document}